\documentclass[12pt]{JHEP3}

\usepackage{amsmath}
\usepackage{epsfig}
\usepackage{amstext,amssymb,amsfonts}

\def\Tr{\,{\rm Tr}\, }

\def\be{\begin{equation}}
\def\ee{\end{equation}}
\def\ba{\begin{eqnarray}}
\def\ea{\end{eqnarray}}

\renewcommand{\H}{{\cal H}}

\newcommand{\W}{{\cal W}}
\newcommand{\Con}{\rm Con}

\title{Limits of minimal models and continuous orbifolds}

\author{
Matthias R.\ Gaberdiel$^a$ and Paulina Suchanek$^{a,b}$ \\ \\
$^a$Institut f\"ur Theoretische Physik, ETH Zurich, \\
$\;$CH-8093 Z\"urich, Switzerland 

\vspace{5mm}
$^b$Institute for Theoretical Physics, University of Wroc{\l}aw, \\
$\;$pl.~M.~Borna 9, 50-204 Wroc{\l}aw, Poland
\\ \\
\email{gaberdiel@itp.phys.ethz.ch, suchanek@itp.phys.ethz.ch}}

\abstract{The $\lambda=0$ 't~Hooft limit of the 2d ${\cal W}_N$ minimal
models is shown to be equivalent to the singlet sector of a free boson theory, thus paralleling 
exactly the structure of the free theory in the Klebanov-Polyakov proposal. In 2d, the singlet 
sector does not describe a consistent theory by itself since the corresponding partition function
is not modular invariant. However, it can be interpreted as the untwisted sector
of a continuous orbifold, and this point of view suggests that it can be made consistent by adding in 
the appropriate twisted sectors. We show that these twisted sectors account for the
`light states' that were not included in the original 't~Hooft limit. We also show that, for the Virasoro
minimal models ($N=2$), the twisted sector of our orbifold agrees precisely with the limit theory of 
Runkel \& Watts. In particular, this implies that our construction satisfies crossing symmetry.}

\begin{document}

\section{Introduction}

Simplified versions of the AdS/CFT correspondence hold the promise of offering
insights into the mechanism that underly the duality. For example, the large $N$ limit of 
the CFT  at weak coupling 
\cite{Sundborg:2000wp,Witten,Mikhailov:2002bp,Sezgin:2002rt} is believed to be dual to a higher spin 
theory on the AdS background \cite{Vasiliev:2003ev} (see for example 
\cite{Vasiliev:1999ba,Bekaert:2005vh,Iazeolla:2008bp,Campoleoni:2009je} for reviews).
Higher spin theories lie in complexity somewhere between field theories and string theories 
in that they contain infinitely many fields, but far fewer than a fully fledged string theory. The corresponding
duality is therefore much more tractable than the  stringy AdS/CFT correspondence, yet contains
sufficiently  much structure in order to capture many of the essential features.
\smallskip

Some years ago, Klebanov \& Polyakov made a concrete proposal along these lines \cite{Klebanov:2002ja}
(for related work see also \cite{Sezgin:2002rt,Sezgin:2003pt}). They conjectured that 
Vasiliev's higher spin theory on AdS$_4$ is dual to the singlet sector of the 3d ${\rm O}(N)$ vector model in 
the large $N$ limit. Recently, impressive evidence in favour of this proposal has been 
found \cite{Giombi:2009wh,Giombi:2010vg,Giombi:2011ya}, see also 
\cite{Koch:2010cy,Douglas:2010rc,Shenker:2011zf,Aharony:2011jz,Giombi:2011kc,MZ} for 
related work.
Last year,  a similar duality was proposed in one dimension less \cite{Gaberdiel:2010pz}: it 
relates a family of higher spin theories on AdS$_3$ \cite{Prokushkin:1998bq,Prokushkin:1998vn} 
to the large $N$ limit of the ${\cal W}_N$ minimal models in 2d (see \cite{Bouwknegt:1992wg} for
a review of ${\cal W}$-algebras). This proposal was motivated by the analysis of the asymptotic symmetries
of higher spin theories on AdS$_3$ \cite{Henneaux:2010xg,Campoleoni:2010zq}, following 
\cite{Brown:1986nw}, see also \cite{Gaberdiel:2011wb,Campoleoni:2011hg} for subsequent work.
By now it has been shown that the spectra of the two theories agree in the $N\rightarrow \infty$ limit 
\cite{Gaberdiel:2011zw}  (see also \cite{Gaberdiel:2010ar}), and correlation functions have been found to 
match \cite{Chang:2011mz,Ahn:2011by,Ammon:2011ua} (see also \cite{Papadodimas:2011pf}).
Generalisations for orthogonal groups have been 
studied \cite{Ahn:2011pv,Gaberdiel:2011nt}, and black hole solutions have been analysed
\cite{Gutperle:2011kf,Ammon:2011nk}; their entropy has (for $\lambda=0,1$) 
been matched to that of the dual CFT \cite{Kraus:2011ds}.

While the proposal of \cite{Gaberdiel:2010pz} is in many ways the natural lower dimensional
analogue of the Klebanov \& Polyakov proposal, the details appear to be somewhat different. 
For the case of the ${\rm O}(N)$ vector model in 3d, there are two conformal fixed points, the
free and the interacting theory, that are believed to be dual to two different higher spin
theories on AdS$_4$. In the lower dimensional version, on the other hand, the ${\cal W}_N$
models possess a line of conformal fixed points in the large $N$ limit that is parametrised
by a 't~Hooft like coupling $0\leq \lambda \leq 1$; this is mirrored by the fact that there exists
a one-parameter family of higher spin theories on AdS$_3$. It seems natural to think of the
theory at $\lambda=0$ as corresponding to the `free' fixed point, and in this paper we make this
correspondence more explicit. The $\lambda=0$ theory corresponds to taking the level $k$ of 
the ${\cal W}_N$ minimal model to infinity, before taking $N\rightarrow \infty$. Working at 
arbitrary finite $N$, we show that the $k\rightarrow \infty$ limit of a ${\cal W}_N$ minimal model,
constructed following \cite{Gaberdiel:2010pz}, 
can be described as the singlet sector of a free theory (consisting of $N-1$ free bosons). 
This is therefore the direct analogue of the Klebanov-Polyakov proposal in one higher
dimension. For $N=2$, the $k\rightarrow \infty$ limit corresponds to taking the $c\rightarrow 1$ limit of 
the Virasoro minimal models, and the limit of \cite{Gaberdiel:2010pz} is analogous to what was 
considered in \cite{Roggenkamp:2003qp} (except that we restrict ourselves to a subset of their
spectrum for which the partition function  converges). 

The resulting conformal field theory is well-defined on the sphere, but it is not modular invariant
because of the singlet constraint, and hence the resulting conformal field theory
is not fully consistent.\footnote{Note that at finite $N$, the central charge equals $c=N-1$
in this limit, and hence the requirement of modular invariance can be clearly posed.}
However, there is a very natural way in which to repair this: we can think of the
singlet condition as an orbifold projection, for which  the above singlet sector is 
the untwisted sector. Then in order to make the theory consistent, all we have to do 
is to add in the twisted sectors. While this sounds straightforward in principle, there
is one somewhat unusual feature: the singlet constraint requires that we orbifold by a continuous compact
Lie group (rather than a discrete group), and thus the analysis requires some care. In particular,
the twisted sectors are labelled by a continuous parameter (that describes the different conjugacy classes of
the orbifold). As we shall see, the ground states of these twisted sectors then have a natural
interpretation in terms of the $k\rightarrow\infty$ limit of the ${\cal W}_N$ minimal models: they
describe  the `light states' that were not considered in the limit of \cite{Gaberdiel:2010pz} 
since they correspond to states where the size of the Young tableaux scales with $k$ (or $N$). 
These light states do not contribute in intermediate channels to the correlators of the 
usual perturbative states from the
untwisted sector, because the 
fusion of states with finitely many Young boxes does not give rise to states where the number of
boxes grows with $k$. 
\smallskip

Given that our orbifold construction is somewhat unusual --- it is the orbifold of $N-1$
free bosons by the continuous group
${\rm SU}(N)/\mathbb{Z}_N$ --- one may worry whether it is in fact consistent. While we 
cannot prove this in general, we can relate our construction for $N=2$ to a theory that is 
believed to be consistent. 
As was mentioned above, the untwisted sector of the $N=2$ orbifold
can be thought of as a subsector of the $c\rightarrow 1$ limit of Virasoro minimal models of 
\cite{Roggenkamp:2003qp}. It also turns out that the twisted sector has a very natural
interpretation: it seems to agree precisely with the alternative $c\rightarrow 1$ limit of the minimal
models that was proposed in \cite{Runkel:2001ng}. In particular, we can show that the spectra
coincide, that the fusion rules of \cite{Runkel:2001ng} are reproduced from our orbifold point of view,
and that the boundary conditions from which the construction of \cite{Runkel:2001ng} 
originated agree with the usual fractional branes of our orbifold theory. (The non-fractional
branes also have a nice interpretation: they correspond precisely to the additional boundary conditions
that were later found in \cite{Fredenhagen:2004cj}.) On the other hand, the limit theory of 
\cite{Runkel:2001ng}  is believed to be consistent --- it appears to coincide with the $c\rightarrow 1$
limit of Liouville theory \cite{Schomerus:2003vv} --- and it has been checked to satisfy crossing 
symmetry, which is a highly non-trivial constraint.\footnote{One may ask why the twisted
sector of an orbifold should by itself satisfy crossing symmetry. The reason is that the contribution
from the untwisted sector in intermediate channels is of measure zero and hence does not modify
the crossing symmetry analysis.}  Since we can relate our construction to a seemingly consistent
conformal field theory, this gives strong evidence in favour of the assertion that our continuous 
orbifold construction makes sense.
\medskip

The paper is organised as follows. In section~2 we explain why the $\lambda=0$ theory can
be described as the singlet sector of a free theory. In section~3 we show that this projection
can be realised as a continuous orbifold, and construct the twisted sector explicitly for 
the case of $N=2$. In section~4 we explain the close connection between the twisted
sector for $N=2$ and the construction of Runkel \& Watts \cite{Runkel:2001ng}. 
Section~5 explains the relation between the twisted sector ground states and the 
`light states' of the ${\cal W}_N$ minimal models for large $k$, and section~6 contains our
summary and some open problems. There are two appendices where some of the 
more technical calculations are described.

\section{Limits of minimal models}

The minimal models we are interested in are the ${\cal W}_N$  coset models 
\be\label{cosets}
\frac{ {\mathfrak su}(N)_k \oplus {\mathfrak su}(N)_1}{{\mathfrak su}(N)_{k+1}} 
\ee
that appear in the proposal of \cite{Gaberdiel:2010pz}. 
The 't~Hooft parameter is defined to be 
\be
\lambda = \frac{N}{k+N} \ ,
\ee
and the limit of \cite{Gaberdiel:2010pz} consists of taking $N,k$ to infinity while keeping 
$\lambda$ fixed. The `free' theory should correspond to $\lambda=0$, i.e.\ to the limit
where we first take $k\rightarrow \infty$, and then $N\rightarrow \infty$.  In this paper we 
shall mostly study the case of finite $N$; in order to relate our analysis to the $\lambda=0$
case of \cite{Gaberdiel:2010pz} we should subsequently take $N\rightarrow \infty$. 
\smallskip

The central charge of the minimal model (\ref{cosets}) equals 
\be
c = (N-1) \Bigl( 1 - \frac{N(N+1)}{(N+k)(N+k+1)} \Bigr) \ ,
\ee
and hence approaches $c\rightarrow N-1$ in the limit $k\rightarrow \infty$. 
There are different ways in which one may take this limit.
In this section we shall define the limit representations by keeping the representation labels 
of ${\mathfrak su}(N)$ fixed while taking the limit; this is the analogue of what was done in 
\cite{Gaberdiel:2010pz}. 

In order to understand the resulting representations in detail, it is convenient to describe the
coset theory in terms of a Drinfeld-Sokolov (DS) reduction. From this perspective, the representations 
of the coset theory are labelled by (see for example \cite{Bouwknegt:1992wg} for an introduction to
these matters) 
\be
\Lambda = \alpha_+\Lambda_+ + \alpha_-\Lambda_- \ ,
\ee
where
\be
\alpha_+\alpha_- = -1 \  , \qquad  \alpha_- = - \sqrt{k_{\rm DS}+N}\ , \qquad 
\alpha_0 = \alpha_+ + \alpha_- \  ,
\ee
and $k_{\rm DS}$ is the level of the DS-reduction; this is 
related to the level $k$ in the coset description via
\be\label{krel}
\frac{1}{k+N} = \frac{1}{k_{\rm DS}+N} - 1  \ .
\ee
Furthermore, $\Lambda_+$ and $\Lambda_-$ are representations of ${\mathfrak su}(N)$. In the limit
$k\rightarrow \infty$, the level of the DS reduction goes to $k_{\rm DS}\rightarrow - N + 1$, and hence
\be\label{2.7}
\alpha_+ \cong   1\ ,    \qquad
\alpha_- \cong - 1 \ , \qquad 
\alpha_0 \cong 0 \ . 
\ee
The eigenvalues of the highest weight state $(\Lambda_+;\Lambda_-)$ with respect to the zero mode of the (non-primary)
spin $s$ fields are (see \cite[eq.\ (6.50)]{Bouwknegt:1992wg})
\begin{eqnarray}\label{uvalue}
u_s(\Lambda) & = &  (-1)^{s-1}\sum_{i_1 <\dots < i_s} \prod_{j=1}^s 
[(\Lambda, \varepsilon_{i_j}) + (s-j)\alpha_0]  \nonumber \\
& \cong &  (-1)^{s-1}\sum_{i_1 <\dots < i_s} 
\prod_{j=1}^s (\Lambda_+-\Lambda_-, \varepsilon_{i_j})    \ ,
\end{eqnarray}
where $\varepsilon_i$ are the weights in the fundamental (vector) representation of 
$\mathfrak{su}(N)$, 
and in going to the second line we have set $\alpha_0 \cong 0$, as follows from (\ref{2.7}).
Thus in the limit $k\rightarrow \infty$, the coset representation $(\Lambda_+;\Lambda_-)$ only depends
on $(\Lambda_+-\Lambda_-)$; for example, for $N=2$, this is just the familiar statement that, 
as $k\rightarrow \infty$, 
\be\label{Kac}
h(r;s) \simeq \frac{(r-s)^2}{4}  \ , 
\ee
where $(r;s)$ are the usual Kac labels.

The irreducible degenerate $\W_N$ representations at $c=N-1$ are thus 
already accounted for by the representations labelled by $(\Lambda;0)$, where $\Lambda$ is an arbitrary 
weight of $\mathfrak{su}(N)$; any other degenerate representation, i.e.\ any representation labelled by
$(\Lambda_+;\Lambda_-)$,  is (at $\lambda=0$) isomorphic to a direct sum of these
\cite{Gaberdiel:2011zw} (see also \cite[Remark 4.1.7]{Roggenkamp:2003qp} for the Virasoro 
case).
In order to determine the actual decomposition, recall that the 
character of the $(\Lambda_+;\Lambda_-)$ representation equals the branching function of the 
level $k=1$ affine character with respect to the finite dimensional ${\mathfrak su}(N)$ representation 
$(\Lambda_+\otimes \Lambda_-^\ast)$, see
\cite{Bouwknegt:1992wg,Gaberdiel:2010pz}. Thus we conclude that the decomposition is 
\be\label{iden}
(\Lambda_+;\Lambda_-) \cong \bigoplus_{\Lambda} N_{\Lambda_+,\Lambda_-^\ast}^{\Lambda} \,
(\Lambda;0) \qquad
\hbox{for $k\rightarrow \infty$ ,}
\ee
where $N_{\Lambda_+,\Lambda_-^\ast}^{\Lambda}$ are the Clebsch-Gordon coefficients
\be
\Lambda_1 \otimes \Lambda_2 = \bigoplus_{\Lambda} N_{\Lambda_1,\Lambda_2}^{\Lambda} \Lambda \ .
\ee
Note that this implies in particular that we have the equivalences 
\be\label{idenf}
({\rm f};0) \cong (0;\bar{\rm f})  \qquad \hbox{and}  \qquad 
(\bar{\rm f};0) \cong (0;{\rm f}) \qquad \hbox{as $k\rightarrow \infty$,}
\ee
where ${\rm f}$ and $\bar{\rm f}$ denote the fundamental and anti-fundamental
representations of $\mathfrak{su}(N)$, respectively.
The natural `charge-conjugation' theory that contains each of these degenerate representations
once is then 
\be\label{8}
\H_{\rm U} = \bigoplus_{\Lambda} \H_{(\Lambda;0)} \otimes \bar{\H}_{(\Lambda^\ast;0)} \ ,
\ee
where the sum runs over all representations of ${\mathfrak su}(N)$, and $\Lambda^\ast$ is
the conjugate representation to $\Lambda$.

\subsection{The dual gravity perspective}

The equivalence of  conformal field theory representations  described by (\ref{iden}) 
(and in particular by (\ref{idenf}))
is also mirrored in the dual higher
spin gravity theory, at least if we subsequently take $N\rightarrow \infty$. Recall from 
\cite{Gaberdiel:2010pz} that the two complex scalar fields labelled by 
$\left[({\rm f};0), (\bar{\rm f};0)\right]$ and $[(0;{\rm f}),(0;\bar{\rm f})]$  always 
have the same mass, but satisfy in general different boundary conditions
since the conformal weights of the corresponding boundary fields are 
\be
h({\rm f};0) = h(\bar{\rm f};0)  = \frac{1}{2} (1 + \lambda) \ , \qquad
h(0;{\rm f}) =h(0;\bar{\rm f}) = \frac{1}{2} (1 - \lambda) \ .
\ee
In our limit we have $\lambda=0$, and hence the two boundary conditions coincide. Thus the
two complex  scalar fields are indistinguishable, i.e.\ they should describe the `same' field.
It is then natural to consider the subtheory that only contains one of the two complex scalar fields;
this is similar to what was proposed (albeit for general $\lambda$) in \cite{Chang:2011mz}. 
The dual CFT then only has one set of representations, say those of the form $(\Lambda;0)$; its
spectrum is thus precisely equal to that in eq.\ (\ref{8}).

\subsection{Interpretation as a singlet sector}

Next we want to show that (\ref{8}) 
actually has a very natural interpretation as the singlet sector of a theory of $(N-1)$ free
bosons. In order to see this, recall that the ${\mathfrak su}(N)$ level $k=1$ theory can be written in 
terms of $(N-1)$ free bosons compactified on the ${\mathfrak su}(N)$ lattice. Written in terms of the affine level one representations, the free theory has thus the form
\be\label{9}
\H_{\rm free} = \bigoplus_{\mu\in P^+_1} \H^{\hat{\mathfrak su}}_{\mu} \otimes 
\bar{\H}^{\hat{\mathfrak su}}_{\mu^\ast} \ ,
\ee
where  $\H^{\hat{\mathfrak su}}_{\mu}$ denotes the affine representation labelled by  $\mu$, and the sum runs
over all integrable level one representations, i.e.\ those representations where the sum of the Dynkin labels
is at most one. Here $\mu^\ast$ is again the conjugate representation to $\mu$.

The $\W_N$ algebra at $c=N-1$ can be identified with the Casimir subalgebra of the level one
affine algebra \cite{Bouwknegt:1992wg}, i.e.\ $\W_N$ is the commutant of the zero mode
algebra $\mathfrak{su}(N)$ in the vertex operator algebra based on $\hat{\mathfrak{su}}(N)_{1}$. 
Thus any representation 
$\H^{\hat{\mathfrak su}}_{\mu}$ can be decomposed into representations of 
${\mathfrak su}(N) \oplus \W_N$, 
\be\label{10}
\H^{\hat{\mathfrak su}}_{\mu} = \bigoplus_{\Lambda}  \Lambda \otimes \H_{(\Lambda;0)} \ ,
\ee
and the usual Howe-type duality arguments (see e.g.\ \cite{Howe} for the basic idea)
imply that the multiplicity 
space with which $\Lambda$ appears in $\H^{\hat{\mathfrak su}}_{\mu}$ is an irreducible 
representation of $\W_N$; by comparing the character (see above), it is then clear that the 
relevant representation must be the one labelled by $(\Lambda;0)$. 
Note that $\Lambda$ runs over all representations of the (finite) Lie algebra for which
the center acts as in $\mu$,  i.e.\  for which $\Lambda-\mu$ lies in the root 
lattice.

Combining (\ref{9}) and (\ref{10}) we now conclude that the free theory has the structure
\be
\H_{\rm free} = \bigoplus'_{\Lambda_1,\Lambda_2} \, (\Lambda_1 \otimes \Lambda_2^\ast ) \otimes
\bigl(\H_{(\Lambda_1;0)} \otimes \bar\H_{(\Lambda_2^\ast;0)} \bigr) \ ,
\ee
where the sum runs over all representations $\Lambda_j$ of ${\mathfrak su}(N)$, with
the only constraint that $\Lambda_1-\Lambda_2$ lies in the root lattice --- this
is indicated by the prime. Here the space is decomposed with respect to ${\mathfrak su}(N) \oplus \W_N$,
both for left- and right-movers.

It is now immediate that the representation space in (\ref{8}) equals precisely 
\be\label{12}
\H_{\rm U} = \H_{\rm free}^{(0)} \ ,
\ee
where the index $(0)$ means that we restrict ourselves to the subspace of $\H_{\rm free}$ for which
the zero mode action $J^a_0 + \bar{J}^a_0$ is trivial, i.e.\ to the states that are singlets under the diagonal
action of the left- and right-moving zero mode. Indeed, requiring this singlet condition simply
means that we restrict each tensor product $(\Lambda_1\otimes \Lambda_2^\ast)$ to the singlet sector;
the trivial representation is contained in $(\Lambda_1\otimes \Lambda_2^\ast)$ if and only if 
$\Lambda_1\cong \Lambda_2$, and if this is the case, it appears with multiplicity one. 
Thus (\ref{12}) follows from the comparison with (\ref{8}). 
If we specialise to the case $N=2$, the spectrum of $\H_{\rm U}$ is a subsector of the spectrum 
proposed in \cite{Roggenkamp:2003qp}.

\section{The continuous orbifold}

The above singlet condition is very reminiscent of what was proposed by
Klebanov \& Polyakov in the corresponding 3d situation \cite{Klebanov:2002ja}. In the
present context, we know that, by itself, the 
singlet sector is  not  a consistent conformal field theory since the partition function
of $\H_{\rm U}$ is not modular invariant. However, there is a natural way to complete the above theory
to a consistent conformal field theory: we can think of the singlet constraint as the effect of an orbifold
projection, and then the completion just consists of adding in the appropriate twisted sectors. 
There is, however, one subtlety here: the relevant orbifold group is a compact Lie group (rather than
a finite discrete group), and hence the analysis requires some care. On the other hand, since
compact Lie groups behave in many respects very similar to finite discrete groups, it should not be too
surprising that a construction along these lines is possible.

\subsection{The orbifold projection}

The simplest way to describe the singlet condition is via the projection operator
\be
P = \frac{1}{|G|} \, \int_G d\mu(g) \, g \ ,
\ee
where $|G|$ is the total volume of $G$ as measured with respect to the Haar measure $d\mu(g)$. 
The following discussion will be described for an arbitrary Lie group $G$; eventually we shall
apply this to the case where the Lie group is $G={\rm SU}(N)/\mathbb{Z}_N$, and even more specifically
to $G={\rm SU}(2)/\mathbb{Z}_2 \cong {\rm SO}(3)$.\footnote{Note that since the representations in 
$\H_{\rm free}$ are all pairs of representations $(\Lambda_1\otimes \Lambda_2^\ast)$
for which $\Lambda_1 -\Lambda_2$ lies in the root lattice, the center 
$\mathbb{Z}_N$ of ${\rm SU}(N)$ acts trivially, and hence the actual orbifold group is 
$G={\rm SU}(N)/\mathbb{Z}_N$.} The partition function from the untwisted sector is then
\begin{eqnarray}
Z_{\rm U} & = & \frac{1}{|G|} \,  \int_G d\mu(g)  
\Tr_{\H_{\rm free}} \bigl( g\, q^{L_0-\frac{c}{24}} \, \bar{q}^{\bar{L}_0 - \frac{c}{24}} \bigr) \nonumber \\
& = & \frac{1}{|G|} \,  \int_{\mathbb{T}/\W} d\hat{\mu}(h)  
\Tr_{\H_{\rm free}} \bigl( h\, q^{L_0-\frac{c}{24}} \, \bar{q}^{\bar{L}_0 - \frac{c}{24}} \bigr) \ , \label{ZU}
\end{eqnarray}
where $h$ is an element in the Cartan torus $\mathbb{T}$, and $\W$ is the Weyl group of $G$. Here we 
have used that any group element $g\in G$ is conjugate to some element in $\mathbb{T}/\W$, as well as 
the fact that the trace only depends on the conjugacy class ${\Con}_g$ of $g$. Finally, $d\hat{\mu}(h)$ is the 
measure
\be
d\hat{\mu}(h) = {\rm vol} ({\Con}_h) \, d\mu(h) \ .
\ee
The above calculation is illustrated for the case of ${\rm SO}(3)$
in appendix~\ref{A0}, for which $Z_{\rm U}$ turns out to equal, see eq.\ (\ref{psiproj})
\be\label{U2}
Z_{\rm U} =  \sum_{r=0}^{\infty} |\chi_r(q)|^2 \ ,  
\ee
with
\be\label{chirdef0}
\chi_{r}(q) =  \vartheta_r(q) - \vartheta_{r+2}(q) \ , \qquad \hbox{and} \qquad 
\vartheta_r(q) = \frac{q^{\frac{r^2}{4}}}{\eta(q)} \ .
\ee
Since $\chi_r(q)$ is the character of the irreducible $c=1$ Virasoro representation 
labelled by $(r+1;1)$ whose conformal dimension equals $h=\tfrac{r^2}{4}$ in the limit
(see (\ref{Kac})), $Z_{\rm U}$ agrees indeed with the partition function of (\ref{8}).

\subsection{The twisted sector} \label{sec:twisted}

As is familiar from orbifolds of discrete groups, the untwised sector of an orbifold does not
define a consistent conformal field theory by itself since the corresponding partition function is not
modular invariant (and hence the theory cannot be consistently defined on higher genus
surfaces). In order to make the theory consistent we therefore have to add the twisted sectors. 

It follows from general orbifold considerations \cite{Dixon:1985jw,Dixon:1986jc} that the 
twisted sectors are labelled by conjugacy classes of group elements.  For the case at hand, the
twisted sectors are thus labelled by elements $h\in \mathbb{T}/\W$. Each
twisted sector (labelled by $h$) then has to be projected onto the states that are invariant
under the action of the centraliser of $h$ in $G$, 
\be
C_h = \{ g\in G : hg = gh \} \ .
\ee
For a generic element $h\in\mathbb{T}/\W$, the centraliser $C_h$
is just the Cartan torus $C_h = \mathbb{T}$. Thus the actual contribution of the $h$-twisted
sector equals
\be\label{3.7}
Z_{{\rm T}(h)} = \frac{1}{|\mathbb{T}|} \, \int_{\mathbb{T}} d\mu(t)  \Tr_{\H_{h}} \Bigl( t\, q^{L_0-\frac{c}{24}}\, 
\bar{q}^{\bar{L}_0-\frac{c}{24}} \Bigr) \ ,
\ee
where $\H_{h}$ denotes the states in the $h$-twisted sector. 
\medskip

Let us illustrate this for the example of ${\rm SO}(3)$, whose untwisted sector is given in 
(\ref{U2}) and worked out in appendix~\ref{A0}. Using the parametrisation  (\ref{hpar})
we can label the elements of $\mathbb{T}/{\cal W}$ by $h=h(\psi)$, where in ${\rm SO}(3)$ we 
have the identifications $\psi\cong \psi+\pi$ and $\psi\cong \pi-\psi$; denoting the 
representative of $\psi$ with $0\leq \psi \leq \tfrac{\pi}{2}$ by $[\psi]$, the elements of 
$\mathbb{T}/{\cal W}$ can thus be labelled by $\alpha\equiv \tfrac{[\psi]}{\pi} \in [0,\tfrac{1}{2}]$. 

The partition function of the $\alpha$-twisted sector is obtained by applying the $S$-modular 
transformation to the  trace of the untwisted sector with the insertion of $h(\psi(\alpha))$, i.e.\ to 
the integrand of (\ref{psiproj})
\be
Z_{{\rm U}}^{(\alpha)}(\tau) = \sum_{n,w\in\mathbb{Z}} \vartheta_{n+w}(q) \, \vartheta_{n-w}(\bar{q})\, e^{2\pi in\alpha} \ .
\ee
The $S$-modular transformation of $\vartheta_r(q)$ equals
\be
\vartheta_r(\tilde{q}) =\frac{1}{\sqrt{2}} \int_{-\infty}^{\infty} ds\,  e^{\pi i r s} \, \vartheta_s(q) \ ,
\ee
where $\tilde{q} = e^{-2\pi i / \tau}$,
and thus
\begin{eqnarray}
Z_{{\rm U}}^{(\alpha)}(-\tfrac{1}{\tau}) & = & \frac{1}{2} \sum_{n,w\in\mathbb{Z}}
\int_{-\infty}^{\infty} ds \int_{-\infty}^{\infty} d\bar{s}  \, e^{2\pi in\alpha}\,
e^{\pi i n (s+\bar{s})} e^{\pi i w (s-\bar{s})}  \, 
\vartheta_{s}(q) \, \vartheta_{\bar{s}}(\bar{q})\,  \nonumber \\
& = &  \sum_{n\in\mathbb{Z}} \sum_{m\in\mathbb{Z}} 
\int_{-\infty}^{\infty} ds  \, e^{2\pi in\alpha} \, e^{2\pi i n s}\, \vartheta_{s}(q) \, \vartheta_{s+2m}(\bar{q}) \label{T1} \\
& = &    \sum_{m,l\in\mathbb{Z}} 
\vartheta_{-\alpha+l}(q) \, \vartheta_{-\alpha+l+2m}(\bar{q}) =
\sum_{m,\bar{m}\in\mathbb{Z},  m-\bar{m}\in 2 \mathbb{Z} }
\vartheta_{-\alpha+m}(q) \,  \vartheta_{-\alpha+\bar{m}}(\bar{q}) \ . \nonumber
\end{eqnarray}
In the second and third line we have used the identity
\be
\sum_{w\in\mathbb{Z}} e^{i\pi w (s-\bar{s})} =2  \sum_{m\in\mathbb{Z}} \delta(s-\bar{s}+2m) \ .
\ee
Finally, the projection onto the invariant states in the $\alpha$-twisted sector then leads to 
\be\label{T2}
Z_{{\rm T}(\alpha)} =   \sum_{m\in\mathbb{Z}} 
\vartheta_{-\alpha+m}(q) \, \vartheta_{-\alpha+m}(\bar{q}) 
\ee
since the index $-\alpha+m$ and $-(-\alpha+\bar{m})$ can be identified
with the left- and right-moving ${\rm U}(1)$ charge, respectively; this can for example
be deduced from the description of the twisted sector in terms of twisted representations
of the affine algebra $\hat{\mathfrak{su}}(2)$, see appendix~\ref{sec:twisrep} for details. Alternatively,
at least for irrational $\alpha$, this projection can also be obtained by demanding invariance under
the $T:\tau\mapsto \tau+1$ transformation. 

Integrating over the different twist sectors labelled by $\alpha$, the total contribution of the twisted sector is then
\begin{eqnarray}
Z_{\rm T} & = &  \int_0^\frac{1}{2} d\alpha \,  \sum_{m\in\mathbb{Z}} 
\vartheta_{-\alpha+m}(q) \, \vartheta_{-\alpha+m}(\bar{q})     \nonumber \\
& = &  \int_0^{' \infty} dx\, \vartheta_x(q) \vartheta_x(\bar{q}) \ . \label{ZT}
\end{eqnarray}
Strictly speaking the points with $x\in\mathbb{N}$ are excluded from this integral since $\alpha=0$
corresponds to the untwisted sector; this is indicated by the prime in the integral. Our twisted sector 
agrees then precisely with the partition function that was considered by 
Runkel \& Watts \cite{Runkel:2001ng}. We shall elaborate on the precise relation further in 
section~4.
\smallskip

{}From the point of view of our orbifold, (\ref{ZT}) only describes the contribution of the twisted sector.
The total partition function should then be obtained by `adding' to (\ref{ZT}) the contribution from
the untwisted sector (\ref{U2}),  which contains the irreducible Virasoro representations with 
$h=\tfrac{r^2}{4}$, $r\in{\mathbb N}_0$. However, in the 
context of our continuous orbifold we have to be careful how to define this sum 
since the untwisted sector can be thought of as a twisted sector in the limit
of vanishing twist. This suggests that the natural way to include the untwisted sector contribution
is to extend the integral in (\ref{ZT}) to include also the integer points. There is a further subtlety 
in that the Virasoro characters for $h=\tfrac{r^2}{4}$ are not just $\vartheta_r(q)$, but equal 
$\chi_r(q) = \vartheta_r(q) - \vartheta_{r+2}(q)$,  see (\ref{chirdef0}), because of the null-vector
at level $r+1$. However, for the purpose of doing the integral this is immaterial since the integer 
points $x\in\mathbb{N}_0$ are of measure zero. Thus we propose that the full partition function equals
\be
Z_{\rm orb} =  \int_{0}^\infty dx \, \vartheta_x(q) \vartheta_x(\bar{q}) \ ,
\ee
without any restriction on the integral. This is then modular invariant since it equals precisely
one half of the partition function of a single uncompactified free boson
\be
Z_{\rm orb} = \frac{1}{2\, \sqrt{\Im(\tau)}\, \eta(q) \eta(\bar{q})} \ .
\ee
However, as will become clear below, the orbifold theory only shares the partition function with
a free boson theory, but is otherwise very different indeed! This is similar to what happened 
in the construction of Runkel \& Watts \cite{Runkel:2001ng}.

\section{The $c\rightarrow 1$ limit of the Virasoro minimal models}

In the previous section we have proposed that the $k\rightarrow \infty$ limit of 
the coset models (\ref{cosets}) can be described in terms of a continuous orbifold 
of a free boson theory by the compact Lie group $G={\rm SU}(N)/\mathbb{Z}_N$. 
This orbifold construction is somewhat unconventional since the orbifold group in 
question is continuous rather than discrete. One may therefore worry whether the
resulting theory is indeed consistent. As we have seen above, at least for 
the case of $N=2$, the partition function of the orbifold theory is in fact modular
invariant. In this section we want to give further evidence for the consistency of our
orbifold for the case of $N=2$.

As we mentioned before the partition function of the twisted sector of the $N=2$ orbifold theory, 
see (\ref{ZT}), agrees with the spectrum of the Runkel \& Watts limit 
\cite{Runkel:2001ng} of the Virasoro minimal models. In this section, we will argue that this 
correspondence goes beyond just the level of the spectrum. In particular, after explaining
the dictionary between the two descriptions in section~4.1 (see also section~4.3), we show
that the fusion rules of \cite{Runkel:2001ng} have a very natural interpretation from our
orbifold point of view (section~4.2). We shall also construct the boundary conditions  of
\cite{Graham:2001tg} that were the starting point of the Runkel \& Watts analysis as
fractional branes of our orbifold (section~4.4). Since the Runkel \& Watts limit is believed to define a consistent 
theory (that can alternatively be described as the $c\rightarrow 1$ limit of Liouville theory, 
\cite{Schomerus:2003vv,Fredenhagen:2004cj}) this in turn also gives strong support
to our proposal that our orbifold construction leads to a consistent conformal field theory.

\subsection{The identifications}\label{sec:ident}

Let us first explain the relationship between the two descriptions in detail. In the analysis
of Runkel \& Watts \cite{Runkel:2001ng}, the Virasoro primary fields at $c=1$ are labelled by
$x\in\mathbb{R}_+ - \mathbb{N}_0$ with $h_x = \tfrac{x^2}{4}$. In terms of our orbifold
description, the primary $\phi_x$ (as well as its Virasoro descendants) comes from the 
$\alpha$-twisted sector, where 
\be\label{rel0}
\alpha =  \frac{[\psi]}{\pi} = 
\left\{
\begin{array}{cl} 
f_x \quad & \hbox{if $0<f_x\leq \tfrac{1}{2}$} \\
1 - f_x \quad & \hbox{if $\tfrac{1}{2}\leq f_x<1$ .} 
\end{array}
\right.
\ee
Here $[\psi]$ is the representative of $\psi$ with $0 < [\psi]  \leq \tfrac{\pi}{2}$ (see the discussion after 
eq.~(\ref{3.7})), and $f_x$ is the fractional part of $x$, 
\be
x  =   f_x +  \lfloor x \rfloor\ ,
\ee
where $\lfloor x\rfloor$ is the largest  integer less than or equal to $x$, i.e.\
$f_x = x - \lfloor x \rfloor$. Note that a representative for the $\alpha$-twist in (\ref{rel0}) 
is the group element $h(\psi)$ with $\psi = \pi x$ in the parametrisation (\ref{hpar}).
With these identifications the spectra of the two descriptions match precisely. Indeed, it follows from
(\ref{twisdec}) that  the $\alpha$-twisted sector (where $0<\alpha\leq \tfrac{1}{2}$) 
can be decomposed in terms of irreducible Virasoro representations as 
\be
{\cal H}^{(\alpha)} = \bigoplus_{n\in\mathbb{N}_0}
 \Bigl( {\cal H}^{\rm Vir}_{h=\frac{(\alpha+n)^2}{4}} \otimes 
 \bar{ {\cal H}}^{\rm Vir}_{\bar{h}=\frac{(\alpha+n)^2}{4}} 
\, \oplus \, {\cal H}^{\rm Vir}_{h=\frac{(1-\alpha+n)^2}{4}} \otimes  
\bar{{\cal H}}^{\rm Vir}_{\bar{h}=\frac{(1-\alpha+n)^2}{4}}  \Bigr)  \ .
\ee
This then accounts precisely for all $\phi_x$-sectors, given the relation (\ref{rel0}) above.

\subsection{Fusion rules}

Next we want to study the structure of the operator product expansion.
It follows from \cite[eq.\ (9)]{Runkel:2001ng} that the fusion of $\phi_x$ with $\phi_y$ only contains
$\phi_z$ provided that either 
\be\label{c1}
\lfloor x \rfloor + \lfloor y \rfloor + \lfloor z \rfloor \ \hbox{is even} \quad \hbox{and}\quad 
|f_x - f_y | < f_z < \min (f_x+f_y, 2- f_x-f_y) 
\ee
or
\be\label{c2}
\lfloor x \rfloor + \lfloor y \rfloor + \lfloor z \rfloor \ \hbox{is odd} \quad \hbox{and}\quad 
|f_x - f_y | < 1- f_z < \min (f_x+f_y, 2- f_x-f_y) \ .
\ee
We now want to explain how to reproduce this constraint from the orbifold point of view. From 
this perspective, the product of a state in the $\alpha_x$-twisted sector with a state in the
$\alpha_y$-twisted sector can only lead to states in the $\alpha_z$-twisted sector
provided that there are representatives $g_x$, $g_y$ and $g_z$ in the corresponding
conjugacy classes such that \cite{Hamidi:1986vh,Dixon:1986qv,Dijkgraaf:1989hb}
\be
g_z = g_x \cdot g_y \ .
\ee
Next we recall from (\ref{3}) and (\ref{Weyl}) that the group elements in the conjugacy
class of $\alpha_x$ can be taken to have $\chi=\psi_x=\pi x$  (with $\theta=\theta_x$ and $\phi=\phi_x$ arbitrary)
in the parametrisation (\ref{4}).
The product of two group elements with $\chi=\psi_x$ and
$\chi=\psi_y$ is then a group element with $\chi=\psi_z$, where 
\be\label{11}
\cos\psi_z = \cos\psi_x \cos\psi_y - \sin\psi_x \sin\psi_y \Bigl[ 
\cos\theta_x \cos\theta_y + \sin\theta_x \sin\theta_y \cos(\phi_x-\phi_y) \Bigr] \ .
\ee
The expression in brackets is bounded by 
\be
-1 \leq \Bigl[ 
\cos\theta_x \cos\theta_y + \sin\theta_x \sin\theta_y \cos(\phi_x-\phi_y) \Bigr] \leq 1
\ee
and hence
\be\label{91}
\min\Bigl( \cos(\psi_x - \psi_y ), \cos(\psi_x + \psi_y)  \Bigr) \leq \cos\psi_z \leq 
\max\Bigl( \cos(\psi_x - \psi_y ), \cos(\psi_x + \psi_y)  \Bigr) \ .
\ee
The further analysis now depends on the parity of $\lfloor x\rfloor + \lfloor y\rfloor$. 
If $\lfloor x\rfloor + \lfloor y\rfloor$ is {\em even} and working with the representatives $\psi_x=\pi x$ and $\psi_y=\pi y$, then
\begin{eqnarray}\label{P1}
\cos(\psi_x - \psi_y ) &=& \cos\left(|f_x - f_y|\, \pi\right) \\
\cos(\psi_x + \psi_y ) &=&   \cos\left((f_x + f_y)\pi\right) = 
\left\{
\begin{array}{cl} 
\cos\left((f_x + f_y)\, \pi\right) \quad & \qquad \hbox{if $f_x + f_y \leq 1$} \\
\cos\left((2 -f_x - f_y)\, \pi\right)  & \qquad \hbox{if $1<f_x + f_y < 2$ ,} 
\end{array}
\right. \nonumber
\end{eqnarray}
where the arguments on the right hand side are all in the interval $[0,\pi]$, for which the 
cosine is injective. Since we also have with $\psi_z=\pi z$
\be 
\cos(\psi_z) =  \left\{
\begin{array}{cl} 
\cos(f_z\, \pi) \quad & \hbox{if $\lfloor z\rfloor  \in 2\mathbb{N}$} \\
 \cos\left((1-f_z)\pi\right) \quad & \hbox{if $\lfloor z\rfloor \in 2\mathbb{N}+1$} 
\end{array}
\right.
\ee
(\ref{91}) implies for $\lfloor z\rfloor$ even 
\be \label{fz}
|f_x - f_y | \leq f_z \leq \min (f_x+f_y, \ 2- f_x-f_y) 
\ee 
while for $\lfloor z\rfloor$  odd we have instead 
\be \label{1-fz}
|f_x - f_y | \leq 1- f_z \leq \min (f_x+f_y, \ 2- f_x-f_y)  \ .
\ee 
This then reproduces precisely (\ref{c1}) and (\ref{c2}), respectively, except that instead of
the strict inequalities `$<$', (\ref{fz}) and (\ref{1-fz}) involve the non-strict inequalities 
`$\leq$'; this will be commented on in section~4.3 below. The analysis for 
{\em odd} $\lfloor x\rfloor + \lfloor y\rfloor$  is essentially identical. Now the analogue of 
(\ref{P1}) is 
\begin{eqnarray}\label{P2}
\cos(\psi_x - \psi_y ) &=& \cos\left((1-|f_x - f_y|)\, \pi\right) \\
\cos(\psi_x + \psi_y ) &=&   \cos\left(|1-(f_x + f_y)|\, \pi\right) \ ,  \nonumber
\end{eqnarray}
and one obtains (\ref{1-fz}) if $\lfloor z\rfloor$ is even, and (\ref{fz}) if $\lfloor z\rfloor$ is odd.
This then accounts for the remaining cases of (\ref{c1}) and (\ref{c2}), again except for replacing strict
inequalities by non-strict inequalities.

\subsection{The full spectrum}

Recall that the reduced part of the 
Roggenkamp \& Wendland \cite{Roggenkamp:2003qp} spectrum (where we restrict
ourselves to the representations of the form $(r;1)$) describes precisely the untwisted
sector of our orbifold, while the Runkel \& Watts spectrum \cite{Runkel:2001ng}
corresponds to the contribution
from the twisted sector. The untwisted sector is crossing symmetric by itself, but does not 
define a consistent theory since the partition function is not modular invariant. On the 
other hand, the twisted sector is usually, i.e.\ for standard discrete orbifolds, not
consistent by itself since the OPE of two twisted sector states typically also involves untwisted
sector contributions. The situation may be slightly different here, since at least crossing
symmetry is already satisfied by the twisted sector itself, and the partition function is
(at least formally) modular invariant: in both calculations, the contribution from the untwisted
sector is of measure zero and therefore does not modify the answer. However, the
orbifold point of view suggests that the theory {\em can} be (and probably should be)
enlarged to contain both twisted and untwisted sector contributions.

Incidentally, the possibility of extending the theory in this manner was already suggested
in \cite{Runkel:2001ng}. As is explained below eq.\ (6) of that paper, 
one can fairly naturally introduce the identity operator (corresponding to $x=0$)
by the formal limit 
\be
{\bf 1} = \lim_{x\rightarrow 0} \tfrac{1}{x} \phi_x  \ , \label{20}
\ee
and they indicate that similar constructions should also work for any other $x\in\mathbb{N}$. 
In terms of the OPE coefficients, this should then in particular mean that one extends the 
strict inequalities in the fusion rules (\ref{c1}) and (\ref{c2}) to non-strict inequalities.
The resulting extended limit theory should then agree with our continuous orbifold.

\subsection{The fractional branes}

The limit theory of Runkel \& Watts \cite{Runkel:2001ng} was constructed 
so as to be compatible with the 
boundary conditions that had previously been considered in \cite{Graham:2001tg}.
These boundary conditions are labelled by $a\in\mathbb{N}$, and the open string
spectrum between the two boundary conditions $a$ and $b$ equals
\be\label{RWbound}
{\cal H}^{\rm open}_{ab} =  \bigoplus_{r=|a-b|}^{a+b-2} {\cal H}^{\rm Vir}_{h=\frac{r^2}{4}} \ ,
\ee
where the sum over $r$ runs over every other integer, i.e.\ $r$ is even or odd depending on the 
parity of $a+b$. We now want to show that these boundary conditions have a natural
interpretation from our continuous orbifold point of view.
\medskip

In order to describe the boundary conditions of the orbifold theory recall that 
the conformal branes of the `mother theory', the 
$\hat{\mathfrak{su}}(2)$ affine theory at level $k=1$, 
are labelled by group elements $g\in {\rm SU}(2)$ \cite{Gaberdiel:2001xm}, where the corresponding 
boundary state is characterised by the gluing condition
\be\label{gluing}
\bigl(J^a_n - g \bar{J}^a_{-n} g^{-1} \bigr) \, |\! | g \rangle\!\rangle = 0 \ .
\ee
Geometrically, the brane corresponding to $g$ describes a D0-brane sitting at the point $g$ on
the group manifold \cite{Alekseev:1998mc}. Under the diagonal group action of the element 
$h\in {\rm SO}(3)\cong {\rm SU}(2)/\mathbb{Z}_2$, the above boundary state gets mapped to 
\be\label{gact}
h\,  |\! | g \rangle\!\rangle = |\! | h\, g \, h^{-1} \rangle\!\rangle \ ,
\ee
as follows directly from (\ref{gluing}): indeed, $h\,  |\! | g \rangle\!\rangle$ satisfies the gluing condition
\be
\bigl( (h\, J^a_n h^{-1}) - h\, g h^{-1}\, (h \bar{J}^a_{-n} h^{-1})\, h g^{-1} h^{-1} \bigr) \,h\,  |\! | g \rangle\!\rangle 
= 0 \ ,
\ee
and if we redefine the basis of the Lie algebra as $\hat{J}^a_n = h J^a_n h^{-1}$, and similarly for the
right-movers, we reproduce precisely (\ref{gluing}) with $g$ replaced by  $h\, g \, h^{-1}$.

The fixed points of this group action are therefore the branes associated to the identity, $g={\bf 1}$, and 
to the non-trivial element of the center, $g=C$. As is familiar from the general construction of 
D-branes (or boundary conditions) in orbifold theories, see e.g.\ \cite{Douglas:1996sw}, the
corresponding D-brane is then a `fractional brane' that will also couple to the twisted
sectors of the orbifold. The fractional branes are characterised by a (in general projective)
representation $R$ of the orbifold group $G$ 
\cite{Douglas:1998xa,Douglas:1999hq,Gaberdiel:2000fe}; this determines the open string 
spectrum between the boundary conditions labelled by $R$ and $S$ as
\be\label{openproj}
Z_{RS} (q)= 
\frac{1}{|G|} \sum_{g\in G} \Tr_{\H}\Bigl( g \, q^{L_0-\frac{c}{24}} \Bigr) \, \chi_R^\ast(g) \, \chi_S(g) \ , 
\ee
where $\H$ is the open string spectrum of the brane before orbifolding, and 
$\chi_R(g)$ is the group character of $g$ in the representation $R$. Using 
\be
\chi_R^\ast(g) \, \chi_S(g) = \sum_Q N_{SQ}{}^{R} \, \chi_Q^\ast(g) \ ,
\ee
where $N_{SQ}{}^{R}$ are the Clebsch-Gordon coefficients for the decomposition of 
$R^\ast \otimes S$ into the representations $Q^\ast$, we can rewrite (\ref{openproj}) as 
\be\label{openproj1}
Z_{RS} (q) =  \sum_Q N_{SQ}{}^{R} \frac{1}{|G|} \sum_{g\in G} 
\Tr_{\H}\Bigl( g \, q^{L_0-\frac{c}{24}} \Bigr) \, \chi_Q^\ast(g) \ .
\ee
Decomposing the open string spectrum ${\cal H}$ with respect to the action of $G$ 
(as was done in (\ref{10}))
\be
\H = \bigoplus_{S} S \otimes \H^{(S)} \ , \qquad \hbox{so that} \qquad
\Tr_{\H}\Bigl( g \, q^{L_0-\frac{c}{24}} \Bigr) 
= \sum_{S} \chi_S(g)\, \Tr_{\H^{(S)}}\Bigl( q^{L_0-\frac{c}{24}} \Bigr) \ ,
\ee
and using the usual orthogonality relation of group characters
\be
 \frac{1}{|G|} \sum_{g\in G}  \chi^\ast_Q(g)\,  \chi_S(g) = \delta_{QS}
\ee
the open string spectrum in  (\ref{openproj1}) consists then precisely of those
states $\H^{(Q)}$ in $\H$ that transform in the $Q$-representation of the orbifold group
\be
Z_{RS} (q) =  \sum_Q N_{SQ}{}^{R}  \Tr_{\H^{(Q)}}\Bigl( q^{L_0-\frac{c}{24}} \Bigr)  \ .
\ee
\smallskip

Returning to the case at hand, if both branes are associated
to the same fixed point, the relative open string before orbifolding is just the 
vacuum ($j=0$) representation of the $\hat{\mathfrak{su}}(2)$ affine theory at level $k=1$; if the two
branes are at different fixed points (one at $g={\bf 1}$, the other at $g=C$), the open string spectrum
between them consists of the $j=\tfrac{1}{2}$ representation of $\hat{\mathfrak{su}}(2)$. Under the action
of the orbifold group these representations decompose as 
\be\label{suVir}
\H_{j=0}^{\mathfrak{su}(2)} = \bigoplus_{l\in \mathbb{N}_0} D_l \otimes \H^{\rm Vir}_{h=l^2} \ , \qquad
\H_{j=\frac{1}{2}}^{\mathfrak{su}(2)} = \bigoplus_{l\in \mathbb{N}_0 +\frac{1}{2}} 
D_l \otimes \H^{\rm Vir}_{h=l^2} \ ,
\ee
where $D_l$ is the spin $l$ representation of $G={\rm SU}(2)$. Since the projection 
(\ref{openproj1}) picks out the states that transform in the $Q$ representation, the requirement
that the open string spectrum is non-empty demands that $Q$  is half-integer if the two
branes in question sit at different fixed points. Thus there is a selection rule for what
representations of the orbifold arise: if the fractional brane sits at $g={\bf 1}$, say, then 
$R$ must be a conventional representation of the orbifold group ${\rm SO}(3)$, i.e.\ have integer spin, 
while for the brane located at $g=C$, the representation $R$ must be projective, i.e.\ have half-integer spin. 
(A natural interpretation of this is to say that the orbifold has `discrete torsion', and that
the representation of the orbifold group at the non-trivial fixed point is therefore projective
\cite{Douglas:1998xa,Douglas:1999hq,Gaberdiel:2000fe}.)

Let us denote by $|\!| g, R  \rangle\!\rangle$ the fractional brane sitting at the fixed point $g$ 
and being characterised by the representation $R$. Then we propose that the branes of 
\cite{Graham:2001tg} are to be identified with the fractional branes in our orbifold as 
\be\label{idenb}
(a) \Longleftrightarrow \left\{\;\;
\begin{array}{ll}
|\!| {\bf 1}, D_{l(a)}  \rangle\!\rangle \qquad & a\in 2\mathbb{N}-1 \\
|\!| C, D_{l(a)}  \rangle\!\rangle \qquad & a\in 2\mathbb{N}  \ ,
\end{array} \right.
\qquad \hbox{where} \quad l(a) = \frac{a-1}{2} \ .
\ee
With this identification the relative open string spectrum reproduces precisely (\ref{RWbound}).
Indeed, the above arguments imply that the projection picks out those Virasoro representations
from  (\ref{suVir}) that transform as $a\otimes b$, and this is precisely what (\ref{RWbound})
amounts to.
\smallskip

Incidentally, this identification is also compatible with the bulk boundary couplings. It follows from 
\cite[eq.\ (14)]{Runkel:2001ng} that the bulk-boundary coupling of the brane corresponding to $(a)$ 
equals
\be\label{bb}
B(a;x) = \sin(\pi a x) \ ,
\ee
where $x\in \mathbb{R}_+$ labels the different bulk fields of their analysis. In terms 
of our orbifold, $B(a;x)$ should be interpreted as the coefficient with which the above fractional
branes couple to the twisted sectors. At least for the case where the representation $R$ is 
not projective --- the situation is more complicated in the projective case \cite{Gaberdiel:2000fe} --- 
the boundary state of the fractional D-brane sitting at the identity $g={\bf 1}$ is 
schematically (i.e.\ up to normalisations) of the form  
\be\label{fractional}
|\!| {\bf 1}, R  \rangle\!\rangle = |\!| {\bf 1}\rangle\!\rangle + \sum_{\alpha} 
\chi_R(h(\alpha))\, |{\bf 1}\rangle\!\rangle_{\alpha} \ ,
\ee
where $|\!| {\bf 1}\rangle\!\rangle$ is the boundary state of the original theory as in
(\ref{gluing}), while  $|{\bf 1}\rangle\!\rangle_{\alpha}$ is 
the Ishibashi state in the $\alpha$-twisted sector.  Here $ \chi_R(h(\alpha))$ is the character of any 
representative $h(\alpha)$ in the conjugacy class labelled by $\alpha$,
evaluated in the representation $R$. For the case at hand, where
we can take $h(\alpha)$ to lie in the Cartan torus and to correspond to the group element 
(\ref{hpar}) with $\psi=\pi \alpha$, we have 
\be\label{bb1}
\chi_{D_l}(h(\alpha)) = \frac{\sin((2l+1) \pi \alpha)}{\sin(\pi\alpha)} \ . 
\ee
Since $x$ and $\alpha$ are related as in (\ref{rel0}), and since 
$(2l+1)=a$, see (\ref{idenb}), we have
\be
B(a;x) = \sin(\pi x) \, \chi_{D_l}(h(\alpha))  \ .
\ee
Thus the bulk-boundary coupling constants agree up to
the irrelevant normalisation constant  $\sin(\pi x)$ that is independent
of the boundary conditions.

\subsection{The bulk branes}

It was observed in \cite{Fredenhagen:2004cj} that the limit theory of Runkel \& Watts
also possesses another class of boundary conditions that are labelled by $s\in\mathbb{R}$. 
Actually, the self-spectrum of these D-branes only depends on $s$ mod $1$, and it is given
by\footnote{Note that the $p$ parameter of \cite{Fredenhagen:2004cj} is related to the $x$ parameter
of \cite{Runkel:2001ng} as $p=\tfrac{x}{2}$.}
\be\label{Ss}
{\cal S}_s = \left\{ x \in \mathbb{R}_+ : - \min(2 f_s,2-2f_s) <  x <  \min(2 f_s,2-2f_s) \ {\rm mod} \ 2 \right\} \ .
\ee
These branes also have a very natural interpretation
from our orbifold point of view: in addition to the fractional branes that are associated to the 
fixed points of (\ref{gact}), the orbifold theory also possesses `bulk branes' that are simply obtained
as orbifold invariant superpositions of the branes of the mother theory, i.e.\ schematically as 
\be\label{bulkb}
|\!| \psi \rangle\!\rangle = \int_\Gamma d\mu(g)\, \Bigl( 
|\!| g h(\psi) g^{-1} \rangle\!\rangle + |\!| g h(2\pi-\psi) g^{-1} \rangle\!\rangle \Bigr) \ , 
\ee
where $\Gamma$ is the set of group elements (\ref{Gammaset}) parametrised by $\eta$ and
$\varphi$, and $d\mu(g)$ is the restriction of the (suitably rescaled) 
Haar measure to $\Gamma$. Note that the second 
term in (\ref{bulkb}) arises because conjugation by the Weyl group element 
$w\in {\rm SO}(3)$ maps  $\psi$ to $2\pi-\psi$, see (\ref{Weyl}). 

Obviously $|\!| \psi \rangle\!\rangle = |\!| 2\pi - \psi \rangle\!\rangle$, and hence
the above boundary conditions are labelled by $\psi\in[0,\pi]$. As we shall
see below, we can identify $s=\tfrac{\psi}{\pi}$, but this then only accounts for
$s\in [0,1]$. In order to understand the origin of the integer part of $s$, we note that there 
is another (hidden) variable characterising these boundary conditions: the above branes are 
not quite the standard bulk branes since each $|\!| g h(\psi) g^{-1} \rangle\!\rangle$ is actually 
fixed by a one-dimensional subgroup of the orbifold group, namely by $g \mathbb{T} g^{-1}$. Thus we 
must specify in addition a representation of $\mathbb{T}\cong {\rm U}(1)$, i.e.\ an integer. This integer 
then extends $s\in [0,1]$ to $s\in\mathbb{R}$. The integer part of $s$ (i.e.\ this integer)
characterises how the boundary conditions (\ref{bulkb}) couple to the twisted sector of the orbifold; 
however, as will become  clear momentarily, it does not play any significant role for the determination
of the self-spectrum, and hence we will not attempt to work this out in detail. Note that
this mirrors the fact that  ${\cal S}_s$ in (\ref{Ss}) also only depends on $s$ mod $1$.

\smallskip

In order to determine the self-spectrum of these boundary conditions (and hence reproduce
(\ref{Ss})) we recall that the open string spectrum between two boundary states
$|\!| g_1 \rangle\!\rangle$ and $|\!| g_2 \rangle\!\rangle$ is simply equal to the 
$g_1^{-1} g_2$ twisted vacuum representation of $\hat{\mathfrak{su}}(2)_1$, see e.g.\
\cite{Gaberdiel:2001xm}. From the point of view of the Virasoro representation theory,
the relevant open string spectrum is thus
\be\label{open}
{\cal H}^{(\beta)}_0 = \bigoplus_{m\in\mathbb{Z}}\,  {\cal H}^{\rm Vir}_{h=\frac{(2m-\beta)^2}{4}} \ ,
\ee
where $\beta$ is determined by the condition that 
\be\label{betadef}
g_1^{-1} \, g_2 = g \, h(\pi\beta) \, g^{-1} 
\ee 
for some $g$. (This just means that $h(\pi\beta)$ is the element in the Cartan torus that is conjugate to 
$g_1^{-1} g_2$.) 

It is now immediate how to determine the open string spectrum of 
(\ref{bulkb}): the self spectrum of $|\!| \psi =\pi s\rangle\!\rangle$ consists of the
$\beta$-twisted vacuum representation, where $\beta$ is defined by (\ref{betadef}), and 
$g_1$ is either conjugate to 
$h(\pi s)$ or  $h(\pi(2-s))$, and likewise for $g_2$. In addition, if both
$g_1$ and $g_2$ are invariant under the {\em same} ${\rm U}(1)$ subgroup of
${\rm SU}(2)$, the relevant open string spectrum must be projected onto the zero
${\rm U}(1)$ charge sector.\footnote{If the two branes have parameters $s_1$ and $s_2$ with 
$s_1-s_2\in\mathbb{Z}$, the open string spectrum must be 
projected onto the states with ${\rm U}(1)$ charge $s_1-s_2$.}   
However, this projection only applies to a set of measure zero since generically 
$g_1 \mathbb{T} g_1^{-1}$ and $g_2 \mathbb{T} g_2^{-1}$ do not coincide.
For the purpose of finding the continuous part of the spectrum we can therefore
ignore this ${\rm U}(1)$ projection. 

In order to work out the resulting open string spectrum explicitly, we can follow the same arguments as 
in section~4.2, see in particular eq.\ (\ref{11}),  to conclude that $\beta$ must satisfy 
\be
\cos (2 f_s \pi) \leq \cos (\pi \beta) \leq 1 \ .
\ee
(This is the condition irrespective of whether $g_1$ and $g_2$ are conjugate to $h(\pi s)$ or  $h(\pi(2-s))$.) 
Thus we conclude that 
\be
- \min(2f_s, 2-2f_s) \leq \beta \leq  \min(2f_s, 2-2f_s) \qquad \hbox{mod $2$} \ .
\ee
Together with (\ref{open}) this then reproduces precisely (\ref{Ss}), apart from the by now familiar difference 
between strict inequalities and non-strict inequalities.

\section{The twisted sectors from the $\W_N$ coset point of view}

In the previous section we have shown that for the case of $N=2$, our orbifold
theory is very closely related to the construction of Runkel \& Watts \cite{Runkel:2001ng}.
In this section we want to return to the general case. We want to explain that the ground states
of the twisted sectors are directly related to the `light' states of the ${\cal W}_N$ minimal models 
in the $k\rightarrow \infty$ limit \cite{Gaberdiel:2010pz,Gaberdiel:2011zw}.

As was explained in detail in section \ref{sec:ident}, for the case of $N=2$
the label of the twist sectors $\alpha\in [0,\tfrac{1}{2}]$ is related to 
the parameter $x$ of Runkel \& Watts 
\cite{Runkel:2001ng} as in (\ref{rel0}); in particular, for $x\in[0,\tfrac{1}{2}]$ we simply
have $\alpha=x$. On the other hand, it is implicit from the analysis of Runkel \& Watts
\cite{Runkel:2001ng} (see also \cite{Fredenhagen:2004cj}) that we can think of the
fields labelled by $x\in[0,\tfrac{1}{2}]$ as the limit of the $(r;r)$ fields for which $r$
is not kept constant as $p=k+2$ is taken to infinity, but rather scales as $r\sim \alpha p$.
Indeed, the conformal dimension of the $(r;s)$ representation has the expansion
\be
h(r;s) \simeq \frac{(r-s)^2}{4}  + \frac{r^2 -s^2}{4 p} + \frac{s^2-1}{4 p^2}    
+ {\cal O}\Bigl(\frac{1}{p^3} \Bigr) \ . 
\ee
Thus we have for $r=s=\alpha p$ 
\be
\left. h(r;r)\right|_{r=\alpha p} \ 
 \simeq \  \frac{\alpha^2 p^2-1}{4 p^2} \ \simeq \ \frac{\alpha^2}{4} \ .
\ee
We therefore conclude that we can identify the ground states of the twisted sectors of our 
continuous orbifold with the `light' states of the $c\rightarrow 1$ limit of the Virasoro minimal 
models.
\medskip

We now want to argue that a similar relation holds for the $\W_N$ case (see also
\cite{Fredenhagen:2010zh} where some aspects of the Runkel \& Watts analysis have been 
generalised to the $\W_N$ case).  Recall from
\cite{Gaberdiel:2010pz,Gaberdiel:2011zw} that the light states of the $k\rightarrow \infty$ limit of the
$\W_N$ coset theory arise for $\Lambda_+=\Lambda_-=\Lambda$, for which the conformal dimension
is of the form
\be
h(\Lambda;\Lambda) = \frac{1}{2 p (p+1)} \, (\Lambda,\Lambda+2\rho) \ ,
\ee
where $p=k+N$, $\rho$ is the Weyl vector of $\mathfrak{su}(N)$, and 
$(\cdot,\cdot)$ denotes the usual inner product on the weight space. Writing 
$\Lambda$ in terms of Dynkin labels, $\Lambda=[\Lambda_1,\ldots,\Lambda_{N-1}]$, 
we have (see e.g.\ appendix~B.1 of \cite{Gaberdiel:2010pz})
\be
\frac{1}{ 2 p (p+1)} \, (\Lambda,2\rho)  = 
\frac{1}{p (p+1)} \, \sum_{j=1}^{N-1} \Lambda_j \frac{ j (N-j)}{2} \leq
\frac{D(N)}{p(p+1)} \sum_{j=1}^{N-1} \Lambda_j \leq \frac{ D(N)\, k}{p(p+1)} \ ,
\ee
where $D(N)$ is some $N$-dependent constant, and we have used that $\Lambda$
is an integrable weight at level $k$ and hence satisfies $\sum_{j} \Lambda_j\leq k$. 
As we take $k\rightarrow \infty$ for fixed $N$, the right hand side goes to zero. Thus
in this limit  we have (compare also \cite{Fredenhagen:2010zh})
\be\label{tLam}
h(\Lambda;\Lambda) \simeq \frac{1}{2 p (p+1)} \, (\Lambda,\Lambda) 
\simeq \frac{1}{2} \, (\tilde\Lambda,\tilde\Lambda) \ , \qquad \hbox{with} \qquad
\tilde\Lambda = \frac{1}{p} \Lambda \ .
\ee
The `light states' are therefore obtained by scaling the representations 
$\Lambda^{(p)}$ with $p$ such that
$\tilde\Lambda=\tfrac{1}{p} \Lambda^{(p)}$ approaches a constant vector. Since 
each $\Lambda^{(p)}$  must be an integrable weight at level $k=p-N$, 
it follows that $\tilde\Lambda$ must satisfy
\be\label{tLam1}
\sum_{j=1}^{N-1} \tilde\Lambda_j \leq 1 \ ,
\ee
where $\tilde\Lambda=[\tilde\Lambda_1,\ldots,\tilde\Lambda_{N-1}]$ in the usual Dynkin basis.
Furthermore we have $\tilde\Lambda_j \geq 0$. 

As in the Virasoro case above, we now want to identify (a subset of) these $\tilde\Lambda$
with the different 
twists of our continuous orbifold. Recall from the discussion of section \ref{sec:twisted} that the different
twist sectors are labelled by $\alpha$,  where $\alpha$ parametrises the elements in 
$\mathbb{T}/\W$, with $\mathbb{T}$ the Cartan torus and $\W$ 
the Weyl group of ${\rm SU}(N)/\mathbb{Z}_N$.
Using the description in terms of twisted representations as in 
section~\ref{sec:twisrep}, it follows that the conformal dimension of the $\alpha$-twisted sector 
ground state equals (see e.g.\  \cite[eq.\ (4.7)]{Gaberdiel:1995mx})
\be
h(\alpha) = \tfrac{1}{2} \, (\alpha,\alpha) \ ,
\ee
where $\alpha$ is now thought of as a weight, with $(\cdot ,\cdot)$ the natural 
inner product on the weight space. The comparison with (\ref{tLam}) thus suggests that we 
should identify
\be
\tilde\Lambda \ = \ \alpha \ .
\ee
As is shown in appendix~\ref{app:A}, the weights $\tilde\Lambda$ 
satisfying $\tilde\Lambda_j\geq 0$ and (\ref{tLam1}) are  in one-to-one 
correspondence with the weights $\alpha$ parametrising the elements 
in $\hat{\mathbb{T}}/\W$, where
$\hat{\mathbb{T}}$ is the Cartan torus of ${\rm SU}(N)$. For the actual quotient space
$\mathbb{T}/\W$, where $\mathbb{T}$ is the Cartan torus of ${\rm SU}(N)/\mathbb{Z}_N$,
 the weights $\alpha$ have in addition to satisfy (\ref{a_1}) and (\ref{a_2}), which
is the analogue of the constraint $\alpha\leq\tfrac{1}{2}$ (rather than $\alpha\leq 1$) for the case of 
${\rm SO}(3) = {\rm SU}(2)/\mathbb{Z}_2$. This therefore demonstrates that the light states
of small conformal dimension can be identified with the ground states of the twisted sectors.
The remaining light states (as well as some of the states corresponding to the scaled 
representations with $\Lambda_+\neq \Lambda_-$) correspond then to descendants 
in these twisted sectors.

\section{Conclusions}

In this paper we have shown that the $\lambda=0$ 't~Hooft limit of the ${\cal W}_N$ minimal models
\cite{Gaberdiel:2010pz} can be identified with the singlet sector of a free boson 
theory. This is the natural analogue of the free fixed point of the ${\rm O}(N)$ vector model that appeared 
in the duality of Klebanov \& Polyakov in one dimension higher 
\cite{Klebanov:2002ja}. The singlet sector of the free boson theory in 2 dimensions is not a consistent
conformal field theory by itself  since the corresponding partition function is not modular invariant. However,
one can think of it as the untwisted sector of a continuous orbifold. This implies that it
can be made consistent by adding in the appropriate twisted sectors. The relevant twisted
sectors correspond precisely to the `light states' of small conformal dimension; they were not included in the 
limit  of  \cite{Gaberdiel:2010pz}.

Our orbifold construction is somewhat unusual in that the orbifold group is continuous (and compact)
rather than discrete. As a consequence one may be worried about the consistency of the resulting
theory. In order to dispel this suspicion we have shown that for $N=2$, i.e.\ the $c\rightarrow 1$
limit of the Virasoro minimal models, our construction is closely related to the model proposed in 
\cite{Runkel:2001ng}. Given that the latter is known to satisfy a number of non-trivial consistency
conditions (in particular crossing symmetry),  this implies that the same is true for our continuous
orbifold, at least for $N=2$. Recently the analysis of  \cite{Runkel:2001ng} was partially generalised to 
$N>2$ in  \cite{Fredenhagen:2010zh}, where it was argued that the limit theory can be identified
with a Toda field theory, see also \cite{Fredenhagen:2007tk}; it would be interesting to check that 
also these limit theories allow for an orbifold interpretation as argued above.

In the context of the higher spin duality, our analysis gives a nice CFT interpretation to 
the `light states' at $\lambda=0$. One may wonder to which extent this description could also
work for $\lambda>0$. Obviously, for $\lambda>0$, the theory is no longer free, but it would be
interesting to understand whether some aspects of the orbifold description survive when the
coupling is switched on. It would also be interesting to understand the relation of these twisted sectors
to the recent proposal that the light states correspond to conical surpluses \cite{arXiv:1111.3381}.

\section*{Acknowledgments}

We thank Stefan Fredenhagen, Rajesh Gopakumar, Tom Hartman, Emil Martinec, Shiraz Minwalla, 
Mukund Rangamani, Ingo Runkel and Ashoke Sen for useful 
discussions. The work of PS is supported by a Sciex grant of the CRUS, and the work of 
MRG is partially supported by a grant from the Swiss National Science Foundation. MRG
is grateful to the Aspen Center of Physics for hospitality while some of this work was 
being carried out.

\appendix

\section{The case of ${\rm SO}(3)={\rm SU}(2)/\mathbb{Z}_2$}

In this appendix we calculate the partition function $Z_{\rm U}$ (see eq.\ (\ref{ZU})) of the untwisted sector 
explicitly for the case of ${\rm SO}(3) = {\rm SU}(2)/\mathbb{Z}_2$. We also explain how the
corresponding twisted sectors can be described in terms of twisted representations of 
$\hat{\mathfrak{su}}(2)$.

\subsection{The untwisted sector}\label{A0}

Let us parametrise an arbitrary group element in ${\rm SU}(2)$ as (see e.g.\ \cite[eq.\ (2.5)]{David:2009xg}) 
\be\label{4}
g(\chi, \theta,\phi) = \left( \begin{array}{cc}
\cos{\chi} + i\sin{\chi} \cos{\theta} & 
i \sin\chi\, \sin\theta\, e^{i\phi} \\
i\sin\chi\, \sin\theta\, e^{-i\phi}  & 
\cos{\chi} - i\sin{\chi} \cos{\theta}
\end{array} \right) \ ,
\ee
where $\chi,\phi\in [0,2\pi]$, while $\theta\in[0,\tfrac{\pi}{2}]$. In order to describe 
${\rm SO}(3) = {\rm SU}(2)/\mathbb{Z}_2$, we have to 
identify $\chi \cong \chi+\pi$, so that for ${\rm SO}(3)$ we only have 
$\chi\in [0,\pi]$. We take the Cartan torus of ${\rm SU}(2)$ to consist of the group elements of 
the form
\be\label{hpar}
h(\psi) = \left( \begin{matrix} \cos\psi \;\; & i \sin\psi \\ i \sin\psi \;\;& \cos\psi \end{matrix} \right) \ ,
\ee
where $\psi\in [ 0, 2 \pi]$; for ${\rm SO}(3)$, the Cartan torus $\mathbb{T}$ is then of the same form, 
except that $\psi\in [0,\pi]$. For
\be\label{Gammaset}
g(\eta,\varphi) = \frac{1}{\sqrt{2}} \, \left( \begin{matrix} e^{i\eta} \;\; & e^{i\varphi} \\ - e^{-i\varphi} \;\; & e^{-i\eta} 
\end{matrix} \right)
\ee
we find
\begin{eqnarray}
g(\eta,\varphi) \, h(\psi) \, g(\eta,\varphi)^{-1} &= &
\left( \begin{matrix}
\cos\psi + i \sin\psi \cos(\varphi-\eta) \quad &  \sin\psi\, \sin(\varphi-\eta) \, e^{i(\varphi+\eta)} \\
- \sin\psi\, \sin(\varphi-\eta) \, e^{-i(\varphi+\eta)} \quad & \cos\psi - i \sin\psi \cos(\varphi-\eta) 
\end{matrix} \right) \nonumber \\[4pt]
& = &  g\bigl(\psi,\varphi-\eta,\varphi+\eta-\tfrac{\pi}{2}\bigr) \label{3}
\end{eqnarray}
in the notation of (\ref{4}). Thus every group element in ${\rm SU}(2)$ is in the conjugacy class of
an element of the Cartan torus, and similarly for ${\rm SO}(3)$. 
\smallskip

The Weyl group of ${\rm SU}(2)$ is $\mathbb{Z}_2$, and it is generated by
the group element
\be
w = \left( \begin{matrix} 0 \;\; & 1 \\ -1 \quad & 0 \end{matrix} \right) \ ,
\ee
which maps the Cartan torus under conjugation to itself
\be\label{Weyl}
h(\psi) \mapsto w \, h(\psi) \, w^{-1} 
= \left( \begin{matrix} \cos\psi & -i \sin\psi \\ -i \sin\psi & \cos\psi \end{matrix} \right) 
= h(2\pi - \psi) \ .
\ee
For ${\rm SO}(3)$, where $\psi\in[0,\pi]$, the Weyl group then identifies $\psi\cong \pi - \psi$. 
In the following it will be convenient to take $\psi \in \mathbb{R}_+$, and to define
$[\psi]$ to be the representative of $\psi$ (after using the identifications $\psi\cong \psi+\pi$ and
$\psi\cong \pi - \psi$) with $0 < [\psi] \leq \tfrac{\pi}{2}$. We shall usually  parametrise the set 
$\mathbb{T}/\W$ instead of $[\psi]$  by $\alpha\equiv\tfrac{[\psi]}{\pi}\in [0,\tfrac{1}{2}]$. 
\medskip

Using the coordinates in (\ref{4}), the Haar measure on ${\rm SU}(2)$ takes the form
\be
d\mu = \sin^2\chi \, \sin\theta \, d\chi \, d\theta \, d\phi \ ,
\ee 
and thus the volume of ${\rm SO}(3)$ is 
\be
|{\rm SO}(3)| =  \int_0^{\pi} d\chi\, \sin^2\chi \, \int_0^{\frac{\pi}{2}} d\theta\, \sin\theta \, 
\int_0^{2\pi} d\phi =  \pi^2 \ ,
\ee
while the volume of the conjugacy class containing $h(\psi)$ equals
\be
{\rm vol}\bigl({\Con}_{h(\psi)}\bigr) = 2\pi \sin^2(\psi) + 2\pi \sin^2(\pi - \psi)
= 4\pi \sin^2(\psi) \ .
\ee
In order to determine the contribution from the untwisted sector recall that the partition function of 
a free boson at the self-dual radius equals
\be\label{singleboson}
Z_{\rm free} = \frac{1}{\eta\, \bar\eta}\, \sum_{n,w} q^{\;\frac{(n+w)^2}{4}}\, \bar{q}^{\; \frac{(n-w)^2}{4}} \ ,
\ee
where $\eta\equiv \eta(\tau)$ is the Dedekind eta function, and 
$\bar\eta\equiv \eta(\bar{\tau})$, with $q=\exp(2\pi i \tau)$ and 
$\bar{q} = \exp(-2\pi i \bar\tau)$. 
Imposing the projection of (\ref{ZU})  then leads to the untwisted sector partition function
\begin{eqnarray}
Z_{\rm U} & = & \frac{4}{\pi}\, \int_0^{\frac{\pi}{2}} d\psi \, \sin^2(\psi) \, 
 \frac{1}{\eta\, \bar\eta}\, \sum_{n,w\in\mathbb{Z}} q^{\;\frac{(n+w)^2}{4}}\, \bar{q}^{\; \frac{(n-w)^2}{4}} 
 \, e^{2in\psi}  \nonumber \\
 & = & \frac{4}{\pi}\, \int_0^{\frac{\pi}{2}} d\psi \, \sin^2(\psi) \, 
 \frac{1}{\eta\, \bar\eta}\, \sum_{n,w\in\mathbb{Z}} q^{\;\frac{(n+w)^2}{4}}\, \bar{q}^{\; \frac{(n-w)^2}{4}} 
 \, \cos(2n\psi) \label{psiproj} \\
 & = & \frac{1}{\eta\, \bar\eta}\, \Bigl(
 \sum_{w\in\mathbb{Z}} q^{\;\frac{w^2}{4}}\, \bar{q}^{\; \frac{w^2}{4}}  
 - \frac{1}{2} 
  \sum_{w\in\mathbb{Z}} q^{\;\frac{w^2}{4}}\, \bar{q}^{\; \frac{(w+2)^2}{4}}  
 - \frac{1}{2} 
\, \sum_{w\in\mathbb{Z}} q^{\;\frac{(w+2)^2}{4}}\, \bar{q}^{\; \frac{w^2}{4}} \Bigr)  
=  \sum_{r=0}^{\infty} |\chi_r(q)|^2 \ , \nonumber
\end{eqnarray}
where $\chi_{r}(q)$ is defined in (\ref{chirdef0}), and we have used that 
\be
\int_0^{\frac{\pi}{2}} d\psi \, \sin^2(\psi) \, \cos(2n\psi) = \left\{
\begin{array}{cl}
\frac{\pi}{4} & n=0 \\
- \frac{\pi}{8} \;\;& n=\pm\, 1 \\
0 & n\in\mathbb{Z} \backslash\{0,\pm 1\} \ .
\end{array}
\right.
\ee

\subsection{Interpretation in terms of twisted representations}\label{sec:twisrep}

The $\alpha$-twisted sector can also be interpreted in terms of twisted representations
of the affine $\hat{\mathfrak{su}}(2)$ algebra, for a review of twisted representations
see e.g.\  \cite[section 3.5]{DAMTP-86-5}. Recall that the free boson theory 
(\ref{singleboson}) is actually equivalent to the level one affine $\hat{\mathfrak{su}}(2)$ theory.
The twisted sectors are then described by twisted representations of the affine 
$\hat{\mathfrak{su}}(2)$ theory. Since the twists are inner, the corresponding twisted algebras
are all isomorphic to the untwisted algebra. 

In order to explain this in more detail, let us fix conventions for the $\hat{\mathfrak{su}}(2)$ affine algebra
at level $k$. In the Cartan-Weyl basis it is generated by the modes 
\begin{eqnarray}
{}[J^3_m,J^\pm_n] & = & \pm J^\pm_{m+n} \ , \qquad
\qquad [J^3_m,J^3_n] = \tfrac{k}{2} \, m   \, \delta_{m,-n} \\
{}[J^+_m,J^-_n] & = & 2 \, J^3_{m+n} + k \, m \, \delta_{m,-n} \ .
\end{eqnarray}
In addition we have the Virasoro modes $L_m$, whose commutation relations are
\begin{eqnarray}
{}[L_m,L_n] & = & (m-n)\, L_{m+n} + \tfrac{c}{12} \, m \, (m^2-1)\, \delta_{m,-n} \\
{}[L_m, J^a_n] & = & - n \, J^a_{m+n} \ .
\end{eqnarray}
The modes of the $\alpha$-twisted algebra are then of the form $K^3_m$, $K^\pm_s$, where 
$m\in\mathbb{Z}$ while the 
modings of the $K^\pm_s$ generators are $s\in\mathbb{Z} \pm \alpha$, respectively.
Furthermore, we denote by $\hat{L}_m$ the Virasoro modes in the twisted representation. 
These modes satisfy formally the {\em same} commutation relations as the $J^a_m$ and $L_m$, i.e.\
\begin{eqnarray}
{}[K^3_m,K^\pm_s] & = & \pm K^\pm_{m+s} \ , \qquad
\qquad [K^3_m,K^3_n] = \tfrac{k}{2} \, m   \, \delta_{m,-n} \\
{}[K^+_r,K^-_s] & = & 2 \, K^3_{r+s} + k \, r \, \delta_{r,-s} \\
{} [\hat{L}_m,\hat{L}_n] & = & (m-n)\, \hat{L}_{m+n} + \tfrac{c}{12} \, m \, (m^2-1)\, \delta_{m,-n} \\
{}[\hat{L}_m, K^a_p] & = & - p K^a_{m+p} \ .
\end{eqnarray}
The two algebras are isomorphic, the isomorphism being given by 
\begin{eqnarray}
\varphi_\alpha(J^\pm_m) & = & K^\pm_{m\pm \alpha} \\
\varphi_\alpha(J^3_m) & = & K^3_m + \tfrac{\alpha}{2}\, k\, \delta_{m,0} \\
\varphi_\alpha(L_m) & = & \hat{L}_m + \alpha \, K^3_m + \tfrac{k}{4} \alpha^2 \delta_{m,0} \ ,
\end{eqnarray}
as one can easily verify explicitly. The inverse map is then simply 
\begin{eqnarray}
\varphi_\alpha^{-1}(K^\pm_s) & = & J^\pm_{s\mp \alpha} \\
\varphi_\alpha^{-1}(K^3_m) & = & J^3_m - \tfrac{\alpha}{2}\, k\, \delta_{m,0} \\
\varphi_\alpha^{-1}(\hat{L}_m) & = & L_m - \alpha \, J^3_m + \tfrac{k}{4} \alpha^2 \delta_{m,0} \ . \label{A.26}
\end{eqnarray}
With these preparations it is now easy to describe the twisted representations. The untwisted
highest weight representations are labelled by $j=0,\tfrac{1}{2},\ldots,\tfrac{k}{2}$, and they are generated
from a highest weight states satisfying 
\be
J^a_n |j\rangle = 0 \quad (n> 0)  \ , \qquad
J^+_0 |j\rangle = 0 \ , \qquad J^3_0 |j\rangle = j |j\rangle \ ,\qquad
L_0 |j\rangle = \tfrac{j(j+1)}{k+2} |j\rangle \ ,
\ee
by the action of the negative modes. The representation has a singular vector of the form 
\be\label{null}
(J^+_{-1})^{k+1-2j} |j\rangle \cong 0 \ ,
\ee
which generates the full null space. 
The twisted representation acts on the same
vector space, but we describe the action in terms of the $K^a_p$ and $\hat{L}_m$ modes, using
$\varphi_{\alpha}^{-1}$. Since $0 < \alpha \leq \tfrac{1}{2}$ --- in fact $0<\alpha<1$ would suffice --- 
the ground state $|j\rangle$ is still highest weight with respect to the twisted modes as
\begin{eqnarray}
K^+_s |j\rangle & = &  J^+_{s-\alpha} |j\rangle = 0 \qquad \hbox{for} \quad s = m+\alpha > 0 \\
K^-_s |j\rangle & = & J^-_{s+\alpha} |j\rangle = 0 \qquad \hbox{for} \quad s = m- \alpha> 0 \ .
\end{eqnarray} 
However, the $K^3_0$ and $\hat{L}_0$ eigenvalues are now shifted as
\be\label{groundeig}
K^3_0 |j\rangle = (j - \tfrac{k \alpha}{2}) |j\rangle \ , \qquad
\hat{L}_0 |j\rangle = ( \tfrac{j(j+1)}{k+2} - \alpha j + \tfrac{\alpha^2}{4}) |j\rangle  \ .
\ee
For the case of $k=1$ (that is of primary interest to us), the possible values of $j$ are
$j=0$ and $j=\tfrac{1}{2}$. Then the corresponding eigenvalues are 
\begin{eqnarray}
&K^3_0 |0\rangle = - \tfrac{\alpha}{2} |0\rangle  \qquad 
&\hat{L}_0 |0\rangle = \tfrac{\alpha^2}{4} |0\rangle \\
&K^3_0 |\tfrac{1}{2}\rangle =  \tfrac{(1-\alpha)}{2} |\tfrac{1}{2}\rangle  \qquad 
&\hat{L}_0 |\tfrac{1}{2}\rangle = \tfrac{(1-\alpha)^2}{4} |\tfrac{1}{2}\rangle \ .
\end{eqnarray}

Thus we conclude that the conformal dimensions of the $\alpha$-twisted representations are 
$\tfrac{\alpha^2}{4}$, and $\tfrac{(1-\alpha)^2}{4}$, respectively. Since the twisted and
untwisted representations are isomorphic as vector spaces, it is straightforward to determine
the character of the twisted representation from the untwisted character using (\ref{A.26}).
Because of the  free boson realisation of the level one theory, the unspecialised characters 
$\Tr_j(q^{L_0-c/24} y^{J^3_0})$ are 
\be
\chi_0(q,y) = \frac{1}{\eta(q)}\, \sum_{n\in\mathbb{Z}} q^{n^2} y^n \ , \qquad
\chi_{\frac{1}{2}}(q,y) = \frac{1}{\eta(q)}\, \sum_{n\in\mathbb{Z}} q^{(n-\frac{1}{2})^2} y^{n-\frac{1}{2}}\ ,
\ee
and hence the corresponding $\alpha$-twisted characters are 
\begin{eqnarray}\label{twisdec}
\chi_0^{(\alpha)}(q) & = &  \frac{1}{\eta(q)}\, \sum_{n\in\mathbb{Z}} q^{n^2} q^{-n\alpha} q^{\frac{\alpha^2}{4}} 
= \frac{1}{\eta(q)}\, \sum_{n\in\mathbb{Z}}  q^{\frac{(-\alpha+2n)^2}{4}}
= \sum_{n\in\mathbb{Z}} \vartheta_{-\alpha+2n} 
 \\
\chi_{\frac{1}{2}}^{(\alpha)}(q) & = & \frac{1}{\eta(q)}\, \sum_{n\in\mathbb{Z}} 
q^{(n-\frac{1}{2})^2} q^{-\alpha(n-\frac{1}{2})} q^{\frac{\alpha^2}{4}}  
= \frac{1}{\eta(q)}\, \sum_{n\in\mathbb{Z}}  q^{\frac{(-\alpha+2n-1)^2}{4}}
= \sum_{n\in\mathbb{Z}} \vartheta_{-\alpha+2n-1}  \ .\nonumber
\end{eqnarray}
This then matches precisely (\ref{T1}). It is also clear from this analysis that the
${\rm U}(1)$ charge equals $-\tfrac{\alpha}{2}+n$ and $-\tfrac{\alpha}{2}+(n-\tfrac{1}{2})$,
respectively, and thus the projection onto the ${\rm U}(1)$ singlet states for
the left-right spectrum leads precisely to (\ref{T2}).

\section{Identifying twists with weights}\label{app:A}

In this appendix we first want to show that the weights $\tilde\Lambda$ satisfying 
$\tilde\Lambda_j\geq 0$ as well as (\ref{tLam1}) are in one-to-one correspondence
with elements in $\hat{\mathbb{T}}/\W$, where $\hat{\mathbb{T}}$ is the Cartan torus
of ${\rm SU}(N)$, see also \cite{DK}. 
Let $\epsilon_i$, $i=1,\ldots, N$ be the usual orthonormal basis, in terms of which the 
roots of  $\mathfrak{su}(N)$  are described by 
\be
e_{i,j} =\epsilon_i - \epsilon_j \ , \qquad i\neq j \in \{1,\ldots, N\} \ .
\ee
The simple roots can be taken to be $e_i\equiv e_{i,i+1}$, $i=1,\ldots, N-1$, and the 
corresponding fundamental weights are 
\be
\lambda_i = \sum_{j=1}^{i} \epsilon_j 
- \frac{i}{N} \sum_{j=1}^{N} \epsilon_j \ , \qquad i = 1,\ldots, N-1\ .
\ee
In this description the Weyl group $\W$ acts by permuting the basis vector $\epsilon_j$. 
Writing $\tilde\Lambda$ as 
\be
\tilde\Lambda = \sum_{s=1}^{N-1} \tilde\Lambda_s \, \lambda_s 
= \sum_{i=1}^{N} l_i \, \epsilon_i 
\ee
we have 
\be
l_j = \sum_{s=j}^{N-1} \tilde\Lambda_s - \frac{B}{N} \ , \qquad 
B = \sum_{s=1}^{N-1} s \, \tilde\Lambda_s \ .
\ee
By construction we have $\sum_j l_j=0$. Note further that since all $\tilde\Lambda_s\geq 0$ it follows
that 
\be\label{cc1}
l_1 \geq l_2 \geq  \cdots \geq l_N
\ee
and the condition that $\sum_s \tilde\Lambda_s \leq 1$ becomes
\be\label{cc2}
l_1 - l_N \leq 1 \ .
\ee
Because of the ordering (\ref{cc1}) this condition is equivalent to $|l_i - l_j|\leq 1$ 
for all $i,j$. 
\smallskip

We now want to show that the space of all $(l_1,\ldots, l_N)$ satisfying (\ref{cc1}) and
(\ref{cc2}) is in one-to-one correspondence with elements in $\hat{\mathbb{T}}/\W$. First we 
recall that the Cartan torus can be identified with the vector space of `weights'
\be
\alpha = \sum_{j=1}^{N} \alpha_j \epsilon_j  \qquad\hbox{with} \qquad
\sum_{j=1}^N \alpha_j = 0  \ ,
\ee
modulo the addition of roots. Because we are only interested in the quotient by the Weyl group,
we can use the Weyl group action to order the components, i.e.\ we may assume without loss
of generality that 
\be\label{order}
\alpha_1 \geq \alpha_2 \geq \cdots \geq \alpha_N \ .
\ee
Now there are two cases to consider: if $\Delta \equiv \alpha_1 - \alpha_N\leq 1$, i.e.\ 
if all $|\alpha_i-\alpha_j|\leq 1$, we identify 
$\alpha$ directly with $\tilde\Lambda$.  Alternatively, i.e.\ if 
$\Delta \equiv \alpha_1 - \alpha_N > 1$, we subtract from $\alpha$ the root $e_{1N}$, i.e.\
we consider 
\be
\alpha' = \alpha - (\epsilon_1 - \epsilon_N) = \sum_{j=1}^{N} \alpha'_j \epsilon_j = 
(\alpha_1-1) \epsilon_1 + \sum_{j=2}^{N-1} \alpha_j \epsilon_j 
+ (\alpha_N+1) \epsilon_N \ .
\ee
Then we reorder (if necessary) the components of $\alpha'$ so that they satisfy again (\ref{order}). 
If the reordering does not involve either $\alpha'_1$ or $\alpha'_N$, then $\Delta' \leq \Delta - 1$
(if either $\alpha'_1$ or $\alpha'_N$ is not reordered) or $\Delta'=\Delta-2$ (if both are not reordered). 
On the other hand, if both $\alpha'_1$ and $\alpha'_N$ are reordered, then either 
$\alpha'_1=\alpha_N+1$ or $\alpha'_1=\alpha_2\leq \alpha_1$ and either 
$\alpha'_N=\alpha_1-1$ or $\alpha'_N=\alpha_{N-1}\geq \alpha_N$. In any case it then follows that
$\Delta'\leq \Delta$ --- the most subtle case arises for $\alpha'_1=\alpha_N+1$ and $\alpha'_N=\alpha_1-1$
for which
\be
\Delta' = \alpha_N+1 - \alpha_1 + 1 = 2 - (\alpha_1-\alpha_N) < 1 \ .
\ee
Continuing in this manner we can thus find a suitable root $e$ so that $\alpha+e$ 
satisfies $\Delta \leq 1$. (Note that it can happen that in the recursion step the value of 
$\Delta$ does not decrease, $\Delta'=\Delta$, 
but this is only the case if  $\alpha'_1=\alpha_2 =\alpha_1$ and  
$\alpha'_N = \alpha_{N-1} = \alpha_N$. It is then clear that at least after $\tfrac{N}{2}$ iteration
steps, the value of $\Delta$ must strictly decrease. Thus the iterative procedure terminates.)

We conclude that any element in $\hat{\mathbb{T}}/\W$ can be brought into a form
satisfying (\ref{cc1}) and (\ref{cc2}). It is also easy to see (by essentially the same arguments)  
that not two elements of this form (with the exception of some elements with $\alpha_1-\alpha_N=1$) 
can differ by a root. This completes the proof of the first statement.
\smallskip

We are actually interested in the Cartan torus $\mathbb{T}$ of ${\rm SU}(N)/\mathbb{Z}_N$. The generator
of the center $\mathbb{Z}_N$ can be identified with
\be
c_N = \frac{1}{N} \sum_{i=1}^{N-1} i \, e_{i} 
= \frac{1}{N} \sum_{i=1}^{N-1} \epsilon_i - \frac{(N-1)}{N}\, \epsilon_N \ .
\ee
The Cartan torus $\mathbb{T}$ is thus obtained from $\hat{\mathbb{T}}$ upon dividing out the multiples of 
$c_N$, and the quotient space $\mathbb{T}/\W$ is obtained from $\hat{\mathbb{T}}/\W$ by dividing out
the lattice that is generated by the vectors 
\be
c_j = \frac{1}{N} \sum_{i\neq j} \epsilon_i - \frac{(N-1)}{N} \epsilon_j \ , \qquad j=1,\ldots, N \ ,
\ee
i.e.\ by the image vectors of $c_N$ under the Weyl group action. In the quotient space
$\mathbb{T}/\W$ we can therefore reduce the vectors $\alpha$ further to those that
satisfy in addition 
\be\label{a_1}
\alpha_j - \alpha_{j+1} \leq \tfrac{1}{2}  \qquad \hbox{for all} \quad  j=1,\ldots,N-1 \ , 
\ee
as well as 
\be\label{a_2}
\alpha_1 - \alpha_N \leq 1 - \max_i (\alpha_i - \alpha_{i+1}) \ . 
\ee
In order to see that (\ref{a_1}) can be achieved, suppose that 
$\alpha_j - \alpha_{j+1} > \tfrac{1}{2}$ for some $1 \leq j \leq N-1$. (Since $\alpha_1-\alpha_N\leq 1$,
this can happen at most for one $j$.) Then it follows that 
\be \label{a-shifted}
 \alpha'=\alpha + \sum_{i=1}^j c_i
\ee 
 after reordering has the form
\be \label{alpha'}
\alpha'  = (\alpha_{j+1}+\tfrac{j}{N}, \ldots, \alpha_N+\tfrac{j}{N},\alpha_1- 1 +\tfrac{j}{N}, \ldots, 
\alpha_j -1 +\tfrac{j}{N}) \ .
\ee
Since $\alpha'_1 - \alpha'_N = 1 - (\alpha_j - \alpha_{j+1}) < \tfrac{1}{2}<1$, 
the vector $\alpha'$ satisfies then condition (\ref{a_1}), as well as (\ref{cc2}). 

In order to see that we can in addition impose (\ref{a_2}), let $j$ be the value for which 
$\alpha_j - \alpha_{j+1}$ is maximal. If 
$\alpha_1 - \alpha_N >  1-(\alpha_j - \alpha_{j+1})$, we consider
$\alpha' = \alpha+\sum_{i=1}^{j} c_j$ of the form (\ref{alpha'}).
Then the differences $\alpha'_i - \alpha'_{i+1}$ for $i\neq N-j$ agree with the 
differences $\alpha_l - \alpha_{l+1}$ with $l\neq j$, while for $i=N-j$ we now have
\be
\alpha'_{N-j} - \alpha'_{N-j+1} = 
\alpha_N + \tfrac{j}{N} - (\alpha_1 - 1 + \tfrac{j}{N}) = 1 - (\alpha_1-\alpha_N) < \alpha_j - \alpha_{j+1}  \leq \frac{1}{2}
\ee
since $(\alpha_1 - \alpha_N) > 1 - (\alpha_j - \alpha_{j+1}) \geq \tfrac{1}{2}$. 
Because all the differences $\alpha'_i - \alpha'_{i+1}$ are smaller or equal than $ \alpha_j - \alpha_{j+1}$,
 the overall difference $\alpha'_1 - \alpha'_N$ now satisfies the condition (\ref{a_2})
\be
\alpha'_1 - \alpha'_N =  1 - (\alpha_{j} - \alpha_{j+1}) \leq 1-  \max_i (\alpha'_i - \alpha'_{i+1}) \ . 
\ee
\smallskip

We close by noting that the allowed non-trivial weights of the level one algebra are of the form 
$\Lambda^{(j)}_i = \delta_{ij}$ for $j=1,\ldots, N-1$, and hence equal in the orthogonal basis 
\be
\Lambda^{(j)} = - \sum_{i=1}^{j} c_j \ .
\ee
It is then manifest from the above discussion that $\hat{\mathbb{T}}/\W$ 
can be written as the union of $\mathbb{T}/\W$, together with the
shifted weights $\Lambda^{(j)}+\mathbb{T}/\W$. The latter weights 
appear in the twisted version of the level one $\Lambda^{(j)}$ representation
(where we twist again by an element in $\mathbb{T}/\W$). This mirrors
precisely what happened for $N=2$, compare eq.\ (\ref{twisdec}).

\bibliographystyle{JHEP}

\end{document}